# Zero field spin splitting in AlSb/InAs/AlSb quantum wells induced by surface proximity effects

Masaya Nishioka, Bruce A. Gurney and Ernesto E. Marinero
*San Jose Research Center, Hitachi Global Storage Technologies, San Jose, CA 95135, USA*

Francisco Mireles
*Centro de Nanociencias y Nanotecnología, UNAM, Apdo. Postal 2681, Ensenada, Baja California, México*

**Abstract:**

InAs quantum well heterostructures are of considerable interest for mesoscopic device applications such as scanning probe and magnetic recording sensors, which require the channel to be close to the surface. Here we report on magnetotransport measurements of AlSb/InAs/AlSb Hall bars at a shallow depth of 20 nm. Analysis of the observed Shubnikov-de Haas oscillations and modeling show that spin splitting energies in excess of 2.3 meV occur at zero magnetic field. We conclude that the spin-splitting results from the Rashba effect due to the band bending in the quantum well. This is caused by substantial electron transfer from the surface to the quantum well and becomes significant when the quantum well is located near the surface.

Electron transport in AlSb/InAs/AlSb quantum well (QW) heterostructures is attractive for device applications for multiple reasons including the pinning of the InAs surface Fermi level above the conduction band minimum, the small electron effective mass and the large conduction band offset.[1-3] These characteristics make this system useful for the fabrication of nano-scale devices for electronic and sensor device applications including scanning probes and magnetic recording sensors, where proximity of the two dimensional electron gas (2DEG) to the surface is crucial in determining the spatial resolution. Therefore it is desirable to understand the effect of proximity on the potential well and carrier populations. Additionally, semiconductors are now being considered for spintronics applications, so that understanding the behavior of the two spin subbands during current flow is essential.

A complete understanding of transport properties in these AlSb/InAs/AlSb QW heterostructures in terms of spin splitting has not been fully investigated. In 2DEG semiconductor heterostructures, the spin degeneracy can be lifted even in the absence of a magnetic field due to the structure inversion asymmetry (SIA) of the confinement potential through the Rashba effect,[4] which can be tuned by applying a gate voltage.[5,6] This non-degeneracy is known to induce a beating pattern in the Shubnikov-de Haas (SdH) oscillations at low magnetic fields due to the superposition of oscillations of different periods originating from the two spin subbands. Such beating patterns have been observed in several 2DEG systems, for example, GaSb/InAs/GaSb.[7,8] However, the zero-field spin splitting and the observation of beating patterns in AlSb/InAs/AlSb heterostructures differ in published results. In this work, we report on magnetoresistance measurements of AlSb/InAs/AlSb heterostructures in which the InAs 2DEG is located near the surface and show that the spin splitting energy is non-zero due to the Rashba effect on account of proximity of the channel to the surface.

The InAs QW heterostructures were grown by molecular beam epitaxy on GaAs substrates.[9] The 2DEG consisted of AlSb (2 nm)/InAs (12.5 nm)/AlSb (2 nm) layers grown on a thick buffer layer. The wafer was capped with dual layers of $Al_{0.80}Ga_{0.20}Sb$ (14 nm) and InAs (4 nm), which placed the top of the InAs QW channel from the surface at $d$ = 20 nm. To achieve low carrier concentrations, the structures were not intentionally doped. Hall bars were fabricated using an electron beam lithography process combined with ion beam etching. The length and width of the transport channel were 80 μm and 5 μm, respectively. Voltage leads on each side of the channel were 1 μm wide and spaced from each other by 26 μm.

Figure 1(a) shows the sheet carrier density $N_s$ and the mobility as a function of temperature of the Hall devices. It is known that, even in non-intentionally doped InAs QWs, $N_s$ can be relatively large.[1] This is because electrons are easily transferred from donor sites into the exceptionally deep InAs QW. Such donors originate from (1) surface states of the InAs cap layer, (2) the bulk of the AlSb barriers and (3) interface states between AlSb and the InAs channel.[1] $N_s$ in our system was found to be $3.9 \times 10^{11}$ cm$^{-2}$ at 10 K. This is about 30 percent larger than what Nguyen *et al* found in AlSb/InAs/AlSb QWs of comparable thickness where the charge transfer from the surface could be neglected due to the spacing from the surface to the QW of $d$ = 500 nm.[10] We attribute the increase of $N_s$ in our samples to efficient charge transfer from surface states into the QW on account of its proximity ($d$ = 20 nm). The mobility increases with decreasing temperature between 50 and 300 K due to the reduction of phonon scattering. Below 50

K, impurity and interface roughness scattering becomes dominant, therefore, the mobility is almost independent of temperature.[11]

Figure 1(b) provides Hall and longitudinal resistance measurements at 1.8 K. The Quantum Hall effect including Hall plateaus and SdH oscillations are readily observed for applied magnetic fields exceeding 1 T. The Hall plateaus are observed at $\rho_{xy} = h/ne^2$ for the filling factor $n$ = 2, 3, 4, 6, 8 and 10. The existence of the $n$ = 3 Hall plateau is due to Zeeman spin splitting. We note that $g$-factor in InAs is relatively large,[12] therefore, the effect of the Zeeman spin splitting at high magnetic field is evident in the spectrum.

In the low magnetic field regime, we observe a beating node in the SdH oscillations (Fig. 2(a)), although it is not as clear as reported in other systems such as, for example, GaSb/InAs/GaSb QWs.[7,8] As mentioned above, studies of zero field spin splitting in the AlSb/InAs/AlSb system have generated conflicting results. Heida *et al.* observed beating patterns and found a non-zero spin splitting energy, though the Rashba coupling parameter $\alpha_R$ did not depended on the gate voltage, which is not consistent with theory.[13] On the other hand, Brosig *et al.* did not observe beating patterns in dark measurements.[14] However, after illumination, some samples showed beating patterns which was attributed to light-induced spatial inhomogeneity of the carriers. Sadofyev *et al.* also found the beating pattern only during and after illumination, but they claimed that this was not due to the spatial non-uniformity of the carriers.[15] Rowe *et al.* found the beating patterns in high carrier density samples ($N_s$ > 2.4×10$^{12}$ cm$^{-2}$), but concluded that this was not due to the zero field spin splitting but due to the magneto-intersubband scattering.[16]

To determine whether the observed beating node in our data originates from spin splitting due to the Rashba effect, we fitted the longitudinal resistance via the conductivity tensor. The latter was calculated using the model of the magnetoconductance earlier employed to study transport in spin orbit interaction (SOI) systems by Luo *et. al.*[7] The spin-resolved magnetoconductivity tensor of the 2DEG at $T = 0$ K is given by the following equation.[17]

$$\sigma_{xx}^{\pm} = \frac{e^2}{\pi^2 \hbar} \sum_n \left(n \mp \frac{1}{2}\right) \exp\left[-\frac{(E_F - E_n^{\pm})^2}{\Gamma(B)^2}\right] \quad (1)$$

where $E_n^{\pm}$ are the spin dependent Landau level energies with Rashba SOI, $E_F$ is the Fermi energy of the 2DEG and $\Gamma(B)$ is the Landau level broadening. $\Gamma(B)$ is assumed to be spin-independent and equal to $\Gamma_0 + \Gamma_1 \sqrt{B/B_0}$ here, in which $B_0 = 1$ T. Then, the total magnetoresistivity is defined as

$$\rho_{xx} \cong (\sigma_{xx}^{+} + \sigma_{xx}^{-}) \cdot \left(\frac{B}{eN_S}\right)^2 \quad (2)$$

in the limit of high quantizing fields ($(\sigma_{xy}^{\pm})^2 \gg (\sigma_{xx}^{\pm})^2$).

The Landau level energies with Rashba SOI is given by[18]

$$E_n^{\pm} = \hbar\omega_c \left(n + \frac{1}{2} \pm \frac{1}{2}\right) \mp \frac{1}{2}\Delta E_{n,so}^{\pm} \quad (3)$$

in which the spin-splitting $\Delta E_{n,so}^{\pm} = \sqrt{(\hbar\omega_c - \mu_B g^* B)^2 + \hbar\omega_c \Delta_R^2 (n + 1/2 \pm 1/2)/E_F}$. Here, $\omega_c$ is the cyclotron frequency, $\mu_B$ is the Bohr magneton, $g^*$ is the effective Landé factor of the InAs quantum well. The magnetic field is applied along the growth direction. $\Delta_R = 2\alpha_R k_F$ is the Rashba spin-splitting energy (at zero magnetic field) and $k_F$ is the Fermi wave number.

The input parameters for our simulation are the values of $N_s$, $\Delta_R$, $\Gamma_0$, $\Gamma_1$, $g^*$ and $m^*$. In Fig. 2(a) we show the simulation results of the SdH oscillations. The agreement with the experimental result is reasonable. The peak locations and amplitudes of the SdH oscillations are well reproduced by the simulation when the Rashba SOI term is included in the model. The observed node at $B = \sim 1$ T is also reproduced by the simulation. The best fit is achieved with $N_s = 3.65 \times 10^{11}$ cm$^{-2}$, $\Delta_R = 2.37$ meV, $m^* = 0.038 m_0$ ($m_0$ = free electron mass), $\Gamma_0 = 1.1$ meV, $\Gamma_1 = 0.2$ meV and $g^* = -16$. The values of $N_+$ and $N_-$ are calculated to be $1.92 \times 10^{11}$ cm$^{-2}$ and $1.74 \times 10^{11}$ cm$^{-2}$, respectively, using the following equation,

$$\Delta_R = \frac{2\Delta N k_F \hbar^2}{m^*} \sqrt{\frac{\pi}{2(N_S - \Delta N)}} \tag{4}$$

in which $\Delta N = N_+ - N_-$ with $N_s = N_+ - N_-$ and $k_F = \sqrt{2\pi N_S}$.

The total carrier density $N_s$ used in the simulation is consistent with that extracted from the Hall measurement. The fitted value of $\Delta_R = 2.37$ meV is quite similar to that obtained by Heida *et al* (~3 meV) for asymmetric QWs.[13] The effective mass value is within the known values of InAs QW effective mass, $0.0320 m_0 \sim 0.0412 m_0$.[19] The effective Landé factor $g^*$ is comparable with $g = -14$ which has been calculated from magnetoresistance in tilted magnetic field measurements.[12] Therefore, we can state that the obtained parameters from our simulation compare well with published results. Moreover, the carrier densities of two channels derived from fast Fourier transform (FFT) of our experimental SdH oscillations are $2.01 \times 10^{11}$ cm$^{-2}$ and $1.82 \times 10^{11}$ cm$^{-2}$, which yields, using $E_\uparrow - E_\downarrow = 2\pi \Delta N \hbar^2 / m^*$, a spin-splitting of $E_\uparrow - E_\downarrow = 2.46$ meV that is very close to what we obtained from the simulation (Fig. 2(b)). These results support the validity of our model, and, therefore, we can conclude that there is spin imbalance due to the Rashba SOI in our AlSb/InAs/AlSb heterostructures.

We propose that non-zero spin splitting at zero magnetic field due to the Rashba effect is caused by an asymmetric electric potential in the QW. The asymmetric potential is the result of the charge transfer from surface donors to the QW, which becomes important when the 2DEG is located near the surface. The value of $\alpha_R$ arising from such an asymmetric potential can be estimated by the following equation.[18,20]

$$\alpha_R = \frac{\hbar^2 E_p}{6m_0} \langle \Psi(z) | \frac{d}{dz}\left( \frac{1}{E_F - E_{\Gamma_7}(z)} - \frac{1}{E_F - E_{\Gamma_8}(z)} \right) | \Psi(z) \rangle \tag{5}$$

Here, $\Psi(z)$, $E_p$, $E_{\Gamma_7}(z)$ and $E_{\Gamma_8}(z)$ are the wave function, the interband matrix element (Kane parameter) and the energies of the valance band edge for $\Gamma_7$ and $\Gamma_8$ bands, respectively. $z$ is the depth from the heterostructure surface. In order to estimate the value of $\alpha_R$, we derived $\Psi(z)$, $E_{\Gamma_7}(z)$ and $E_{\Gamma_8}(z)$ through a self-consistent (1D Schrödinger-Poisson) calculation of the band structure of the AlSb/InAs/AlSb heterostructure.[21] In Fig. 2(c), we plot $\Psi(z)$, $E_{\Gamma_7}(z)$, $E_{\Gamma_8}(z)$ and $E_{\Gamma_6}(z)$ (the energy of the conduction band edge for $\Gamma_6$ band) as a function of $z$. We found that the calculated non-zero slopes of $E_{\Gamma_7}(z)$, $E_{\Gamma_8}(z)$ and $E_{\Gamma_6}(z)$ in the QW are mainly due to the charge transfer from the capping InAs layer to the QW. Using these parameter values, the value of $\alpha_R$ is estimated to be $5.37 \times 10^{-12}$ eV·m, which is close to the value we obtained from the simulation ($7.82 \times 10^{-12}$ eV·m).[22] We note that the linear Dresselhaus contribution to the spin-splitting in InAs leads to $1.76 \times 10^{-12}$ eV·m while the interface contribution is just $0.3 \times 10^{-12}$ eV·m. Therefore, we conclude that the charge transfer from the surface to the QW is the dominant mechanism that leads to the spin-splitting in our system through a sizable Rashba SOI.

It is noted that in our system neither the carrier non-uniformity nor the magneto-intersubband scattering play an important role. Brosig *et al*. considered the non-uniformity of the carriers as the origin of the beating pattern in their large device (1 mm × 100 μm).[14] However, in our experiments, the magnetoresistance was measured under dark conditions using a significantly smaller device, therefore, the non-uniformity can be neglected. The magneto-intersubband scattering becomes important only when the second subband is occupied. However, the carrier concentration of our sample at 1.8 K is too low to populate the second subband. We also note that the previous studies employed AlSb/InAs/AlSb QWs in which the InAs was located deeper from the surface than in the devices studied in this work.

To summarize, we have fabricated Hall devices in which the InAs QW is located near the surface and have measured their magnetoresistance. From our simulation of the SdH oscillations, we are able to show that there is non-zero spin splitting energy of about 2.4 meV at zero magnetic field which we attribute to the Rashba effect. The origin of the band bending at the quantum well and its concomitant spin-orbit induced spin splitting is likely due to the charge transfer from the surface to the QW, which becomes increasingly important for QWs as they are located near the surface, which is required of nanoscale magnetic field sensors and future magnetic recording readback sensors.

F.M. acknowledges the partial support of DGAPA-UNAM project IN113-807-3. The authors gratefully acknowledge M. Field, A. Ikhlassi, G. Sullivan, and B. Brar of

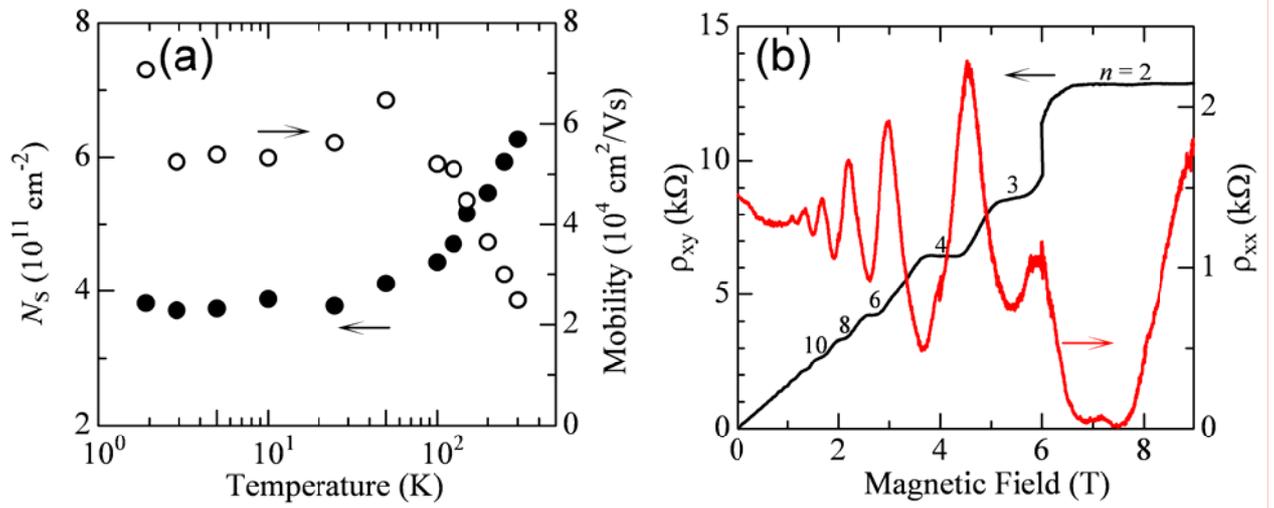

Fig. 1 (a) Temperature dependence of the sheet carrier density (closed circles) and the mobility (open circles). (b) Magnetic field dependence of the longitudinal (red line) and the Hall resistance (black line) at 1.8 K.

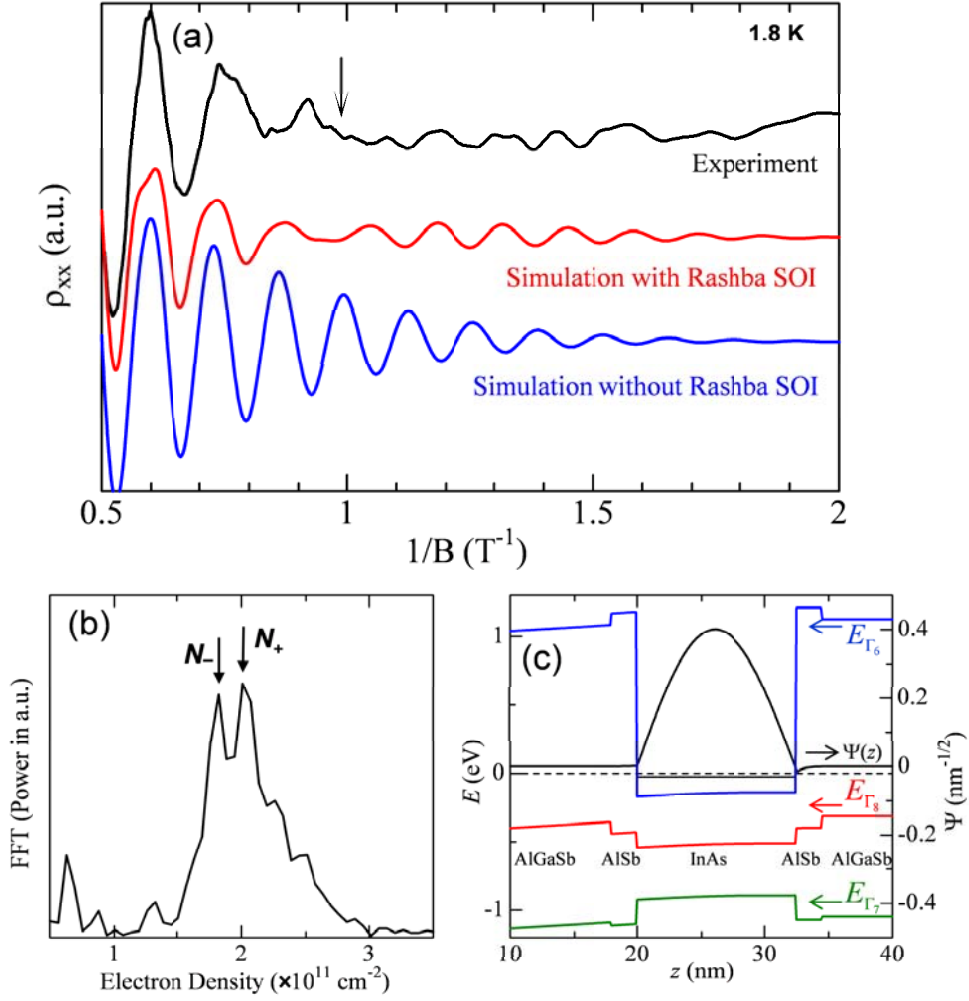

Fig. 2 (a) SdH oscillations at low magnetic field at 1.8 K. The black, red and blue lines show the measured data and the simulation data with and without the Rashba SOI, respectively. The location of the beating node at ~ 1T is indicated by the arrow. The increase of the measured magnetoresistance at $B \sim 0.5$ T is likely due to diffusive boundary scattering. (b) FFT analysis of the SdH oscillations between 0.15 and 4 $T^{-1}$. The arrows indicate the location of two channels. (c) Band structure and wavefunction in the heterostructure obtained by self-consistent Schrödinger-Poisson calculations. The donor concentrations of $3.9 \times 10^{19}$ and $5 \times 10^{14}$ $cm^{-3}$ in the InAs and AlGaSb layers, respectively, and the acceptor concentration of $3.5 \times 10^{16}$ $cm^{-3}$ in the AlSb layers were assumed. The band bending of the bandstructure at the QW is due to charge transfer from the surface. The estimated carrier density at the QW region is $3.82 \times 10^{11}$ cm$^{-2}$ and $E_1 - E_F$ = -24.05 meV. Here, $E_1$ is the first subband energy in QW. The solid horizontal line in QW and the dashed line indicate the location of $E_1$ and $E_F$, respectively.